\documentclass[12pt]{article}
\usepackage{amsmath}%
\usepackage{amsfonts}%
\usepackage{amssymb}%
\usepackage{graphicx}
\begin{document}
\title{Pull over strategy for the evacuation of an emergency vehicle in highways}
\author{Najem Moussa \thanks{e-mail: najemmoussa@yahoo.fr}
\\ D\'{e}partement de Math\'{e}matique et Informatique,
\\ Facult\'{e} des Sciences, B.P. 20 - 24000 - El Jadida, Morocco}
\maketitle
\begin{abstract}
By using a simplified two-lane cellular automata model for traffic
flow, we simulate an evacuation of an emergency vehicle (EV) in
highways. We introduce the pull over strategy to facilitate the
evacuation of the EV in highways. We demonstrate numerically that
this pulling over of cars promotes a faster moving of the EV in
traffic flow and minimize the EV's impact on the overall traffic.
\newline\ Pacs numbers: 45.70.Vn,
02.50.Ey, 05.40.-a
\newline\ \textit{Keywords:} Traffic; Cellular Automata; Emergency Vehicle;
Pull over, Arrival Times
\end{abstract}
Emergency vehicles (EV's) are the vehicles involving in special
events or circumstance in order to provide emergency services to
the public. Police cars, ambulances, fire trucks are some examples
of EV's. During the evacuation, there exists a great task for
reaching the destination (e.g. hospital) as quickly as possible
and with minimal interruptions. Death situation may occur in
emergency evacuation if the arrival time of the EV for reaching
its destination is delayed because of the lack of safe exit
routes. The objective of this paper is to introduce a pull over
strategy in order to facilitate the evacuation of the EV in
highways.
\newline\ Cellular automata (CA) traffic modeling is
always considered as an efficient tool for microscopic traffic
simulations because the CA traffic models are very simple and more
efficient when used in computer simulations (see the review
\cite{chow}). The first CA for one-lane traffic flow is the NaSch
model \cite{ns}. Despite its simplicity, the model is capable of
capturing some essential features observed in realistic traffic
like density waves or spontaneous formation of traffic jams. To
describe more complex situations such as multi-lane traffic,
extensions of the NaSch model have been proposed where additional
rules are added for lane changing cars [2-9].
\newline\ In CA, time and space are discrete. The space is represented as a uniform
lattice of cells with finite number of states, subject to a
uniform set of rules, which drives the behavior of the system.
These rules compute the state of a particular cell as a function
of its previous state and the state of the neighboring cells. The
NaSch model is a one-dimensional probabilistic CA which consists
of $N$ cars moving on a one-dimensional lattice of $L$ cells with
periodic boundary conditions (the number of vehicles is
conserved). Each cell is either empty, or occupied by just one
vehicle with velocity $v=1,2,...,v_{\max }$. We denote by $x_{k}$
and $v_{k}$ the position and the velocity of the {\it kth} car at
time $t$ respectively. The number of empty cells in front of the
{\it kth} car is denoted by $d_{k} =x_{k+1} -x_{k} -1$ and called
hereafter as the gap. Space and time are discrete. At each
discrete time-step $t\rightarrow t+1$ the system update is
performed in parallel for all cars according to the following four
subrules :
\begin{enumerate}
  \item $v_{k}\leftarrow \min \left( v_{k}+1,v_{\max }\right)$
    \item $v_{k}\leftarrow \min \left(v_{k},d_{k}\right)$
   \item $v_{k}\leftarrow \max \left( v_{k}-1,0\right) $ with probability $p$
   \item $x_{k}\leftarrow x_{k}+v_{k}$
\end{enumerate}
Highways are the most frequently used means to evacuate injured
populations. Emergency vehicle is the vehicle, which is given the
highest priority on a road in order to reach its destination as
quickly as possible. When the EV encounters traffic, the other
cars are supposed to pull over to the side of the roadway and give
way to this EV. Thus, the objective of the pull over strategy is
to circulate traffic around the EV to allow it to travel faster
than the normal traffic flow and causing minimal local disturbance
to the freely moving cars.
\newline\ The traffic model we considered here consists of two coupled lanes
where lane changing of vehicles are performed according to some
additional rules. We consider $N+1$ vehicles evolving in the
circuit of size $2*L$. Cars with total number $N$ and one EV are
all allowed to move throughout the lattice. The vehicle movement
is calculated in a two-step process, following \cite{Rick}. First
vehicles change lanes, then they advance. Let us divided the $N+1$
vehicles on two ensembles, one contained $n+1$ vehicles (one EV
and $n$ cars) and the other contained $N-n$ cars. The number $n$
is defined as the number of the leading cars of the EV on the same
lane or on the opposite lane, within a distance range $D_{1}$.
Thus, $n$ may take one of the values {0,1,2}. $D_{1}$ will
represent the interaction horizon of the EV. The above stated
$n+1$ vehicles are subject to the following lane changing rules
which we call hereafter as pull over lane changing. This lane
changing rules are based on the measure of the local densities in
the same and opposite lanes, i.e., the fraction of the $D_{2}$
sites in front of the EV which are occupied. If the EV is on the
lane with low local density, then the EV will not change lanes. In
the contrary case, the EV will attempt to change lanes. Each of
the $n$ cars will attempt to change lane if it is on the lane with
low local density. But, if the car is on the lane with high local
density, it will not change lanes. Neither of the $n+1$ vehicles
will change lanes, in the situation where the local densities in
both lanes are equal. Let's note that these tentative of lane
changing would only be succeeded if the target site is vacant.
Figure 1 illustrates a pull over lane changing pattern of the
$n+1$ vehicles. In order to let the model as simple as possible,
we suppose that $D_{1}=D_{2}/2$.
\newline\ The vehicles which are concerned by the pull over
lane changing rules are the EV and its $n$ leading cars. However,
the others $N-n$ cars are subject to the traditional lane changing
rules \cite{Chow2}:
\begin{enumerate}
  \item $min(v_{k}+1,v_{max})>d_{k}$
    \item $d_{k,other}>d_{k}$ and $d_{k,back}> l_{back}$
   \item $p_{ch}>rand()$
\end{enumerate}
Here $d_{k,other}$ (resp. $d_{k,back}$) denotes the gap on the
target lane in front of (resp. behind) the car that wants to
change lanes. $l_{back}=v^{b}_{o}+1$, where $v^{b}_{o}$ is the
velocity of the following car in the opposite lane. The parameter,
$p_{ch}$ is the lane-changing probability and $rand()$ stands for
a random number between 0 and 1.
\newline\ For comparison, we shall also consider the version of the model where
no pulling over is considered. Let us denote by model A (resp.
model B) the model where the pulling-over strategy is adopted
(resp. not adopted). In model B, all the $N+1$ vehicles (including
the EV) perform lane changing according to the traditional rules,
defined above. In addition, we allow the EV to change lanes
without looking back and to move with higher maximal velocity in
order to reach its destination as quickly as possible. Thus, for
the EV, we set $l_{back}=0$, $p_{ch}=1$ and $v_{max}=7$.
\newline\ The simulation results are depicted in figure 2, where we
show the pull over effect of the general traffic in response to
the presence of an EV. Hence, a considerable improvement of the EV
traffic is obtained when the surrounding cars pull over to the
side of the road. When the EV encounters traffic in its
interaction horizon, it requests its leading car to change lanes,
according to the pull over strategy rules. With success, the EV
can then change lane and accelerate forward until a car is
detected ahead, at which time it requests its new leading car to
change lanes and so on. As figure 2 shows, the EV can reach
velocity greater than the normal traffic, for almost all values of
the global density. Moreover, we observe that the EV mean speed
increases with the interaction horizon $D_{1}$, especially at low
densities.
\newline\ It is clear that the pull over of cars to the side
of the road can reduce, at least locally, the global traffic of
the vehicles. The simulation results show that the pull over
strategy doesn't have any noticeable effect on the global traffic
(Fig. 2) . However, locally, it corresponds to a temporary
reduction in highway capacity along the path of travel of the EV
when cars pull over to the side of the highway. In other words,
the EV behaves like a moving bottleneck which incurs a capacity
reduction all along its route. We believe that the presence of a
great number of EV's in the lattice may effectively induce a
considerable reduction of the mean speed of the vehicles.
\newline\ In figure 3, we show the effect of the interaction
horizon $D_{1}$, on the mean speed of the EV and that of the other
cars. The EV mean speed increases rapidly with $D_{1}$, then
reaches a stationary value. This value corresponds to the maximum
speed that can be provided by the pull over strategy. Hence, more
the EV request is earlier, better is the evacuation. The results
show also that the mean speed of cars undergoes a very weak
reduction when we increase $D_{1}$. Therefore, our results support
well the fact that, the pull over strategy causes minimal local
disturbance to the overall traffic.
\newline\ To get more information on the spatial organization of
the vehicles, one can calculate the distance-headway distribution
of the vehicles, i.e., the instantaneous gap between successive
vehicles. Figure 4a shows the distributions of the distance
headway of the EV in free-flow regime, for the models A and B. In
model B, the distance headway distribution exhibits a single peak
at a distance $D\approx5$, which corresponds to the maximal speed
of the cars. Because the EV have a greater maximal speed, it is
obliged to slow down every time it will be hindered; so as not to
hit its leading cars. The EV behaves therefore like a simple car.
In model A, the distance headway distribution of the EV exhibits
two different peaks, indicating the emergence of two different
behaviors in the evacuation of the EV. The first behavior
corresponds to the situation where the pull over lane changing of
the leading car is not succeeded. The second behavior is when the
leading cars pull over to the side of the road. In this last case,
a larger gap appears in front of the EV, which allow it to
accelerate and to move freely. Our results show also that, as the
interaction horizon increases, the second behavior will dominate
the first one. The direct consequence of this effect, is the
increase of the EV mean speed (see Fig. $3$). Figure 4b
illustrates the effect of the pull over strategy on the distance
headway of the EV when the jammed regime is considered. Here, a
new peak appears in the diagrams at a distance $D=1$. This peak is
a consequence of the confinement of the EV inside the jamming
wave. Besides, thanks to the pull over strategy, the EV can escape
from congested region of cars and win spaces ahead where it can
move freely.
\newline\ The fundamental characteristic in emergency evacuations is the
arrival time of the EV. During the evacuation, there exists a
great task for reaching the destination (e.g. accident site) as
quickly as possible and with minimal interruptions. In figure 5,
we show the distributions of the arrival time of the EV for
different values of the system parameters. In our simulation, we
define the arrival time as the time for the EV to pass on a site
two times in succession. So, we observe that, the model A provides
better arrival times than the model B. Moreover, the arrival time
distribution is shifted towards the lower values when we increase
$D_{1}$. If we compare the arrival time distributions for
different values of the global density, we observe that the
arrival time increases when the global density is increased. Yet,
the congestion is the principal factor which delays the arrival
time of the EV.
\newline\ In summary, we have proposed a simplified CA traffic model
to simulate evacuation of the EV in highways. We introduced the
pull over strategy in order to make easier the evacuation of an
EV. Indeed, we have demonstrated numerically, that thanks to this
strategy, the EV travels faster and then arrives to its
destination with a minimum delays. Besides, the pull over strategy
provokes minimal local disturbance to the overall traffic.
However, in order to make a success of the pull over strategy, it
is important that all the drivers must be very cooperative with
the EV, when they are requested to execute the pull over lane
changing.
\newpage\

\newpage\ \textbf{Figures captions}
\begin{quote}
\textbf{Figure 1}. Pull over lane changing pattern of the $n+1$
vehicles. The colored cars are the leading cars of the EV. Here,
we considered the case where $n=2$. In figure 1a, the local
density of the the red car in its lane (resp. its opposite lane)
is equal to $2/D_{2}$ (resp. $3/D_{2}$). Hence, the red car will
change lanes. For the blue car of figure 1a, the local density in
its lane (resp. its opposite lane) is equal to $3/D_{2}$ (resp.
$2/D_{2}$). Hence, the blue car will not change lanes. The local
density of the the EV in its lane (resp. its opposite lane) is
equal to $2/D_{2}$ (resp. $3/D_{2}$). Hence, the EV will not
change lanes. In figure 1b, the situation is the reverse of what
is in the figure 1a. The red car will not change lanes because it
is on the lane with high local density. The blue car will change
lanes because it is on the lane with low local density. The EV
will change lane because it is on the lane with high local
density.
\newline\ \textbf{Figure 2}. Variation of the mean speed of the EV
($<v>_{EV}$) and that of the other cars ($<v>_{cars}$) with
respect to the variation of the global density, for the models A
and B. All the diagrams of $<v>_{cars}$, corresponding to the
models A and B, merge into one curve.
\newline\ \textbf{Figure 3}. Variation of the mean speed of the EV
($<v>_{EV}$) and that of the other cars ($<v>_{cars}$) with
respect to the variation of the interaction horizon $D_{1}$, for
the model A (a) the global density is equal to 0.05 and (b) the
global density is equal to 0.20 .
\newline\ \textbf{Figure 4}. Probability distribution of the distance headway
of the EV for the models A and B (a) the global density is equal
to 0.05 and (b) the global density is equal to 0.20 .
\newline\ \textbf{Figure 5}. Probability distributions of the arrival
time of the EV for the models A and B (a) the global density is
equal to 0.05 and (b) the global density is equal to 0.20 .
\end{quote}
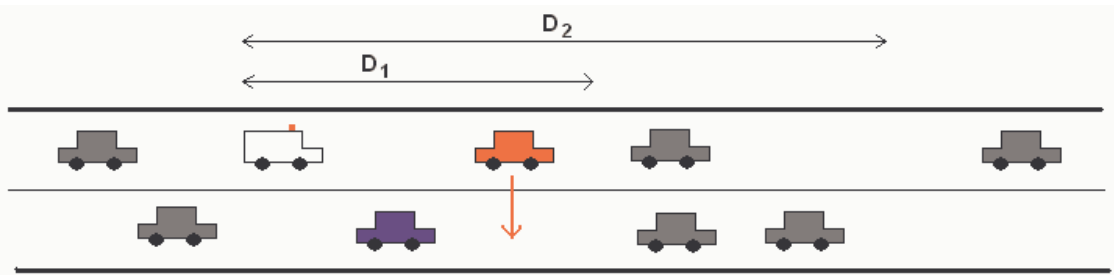
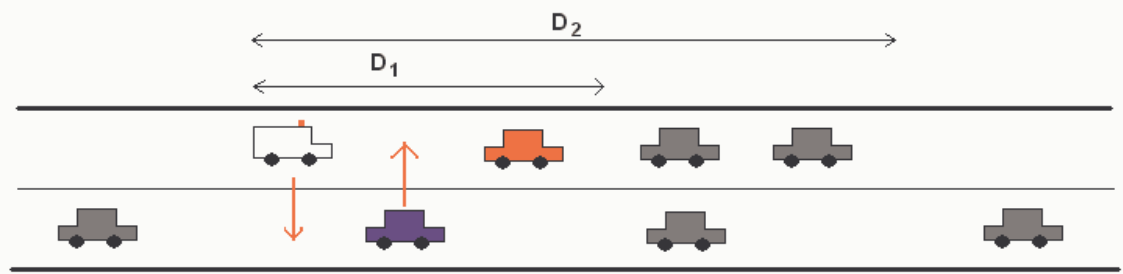
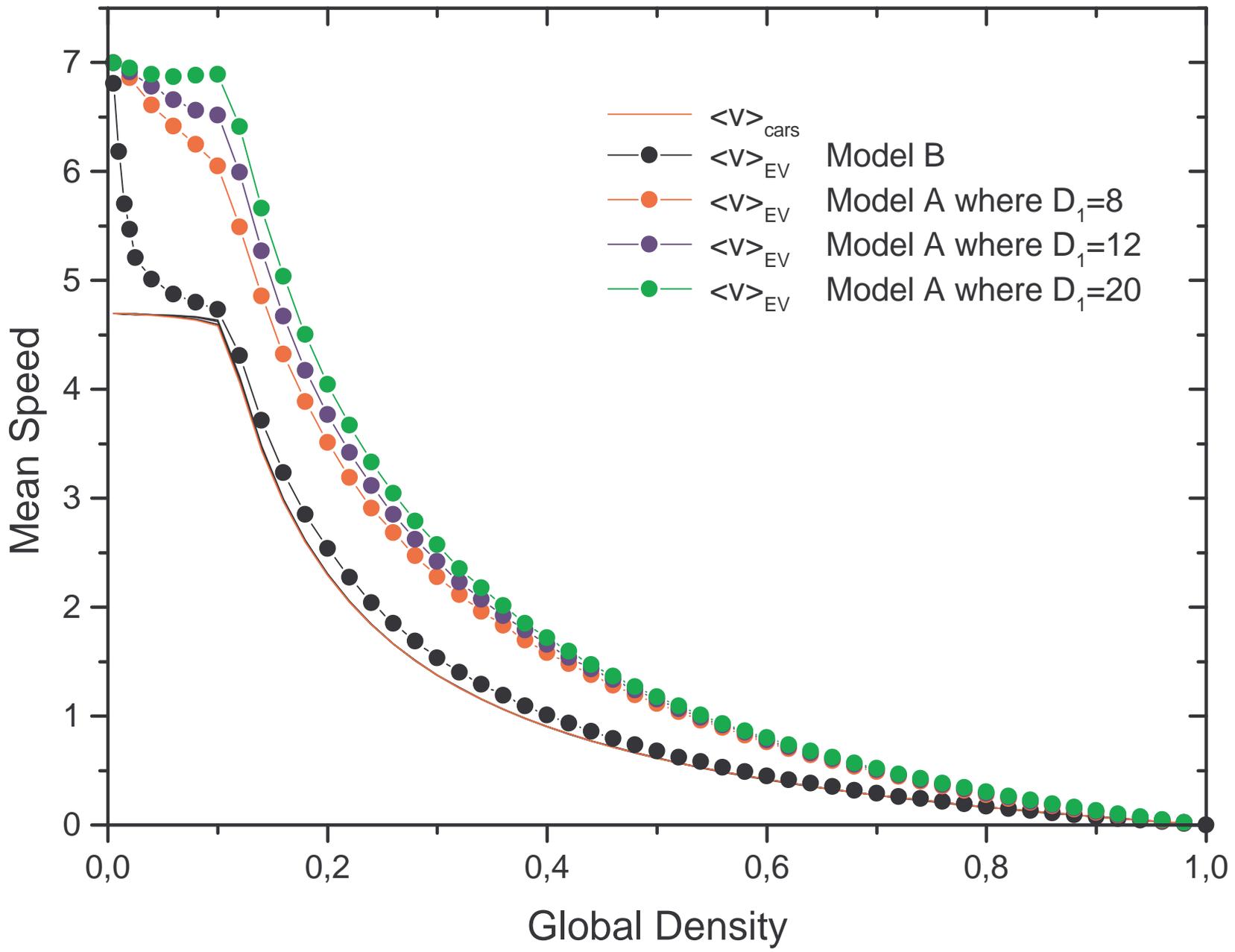
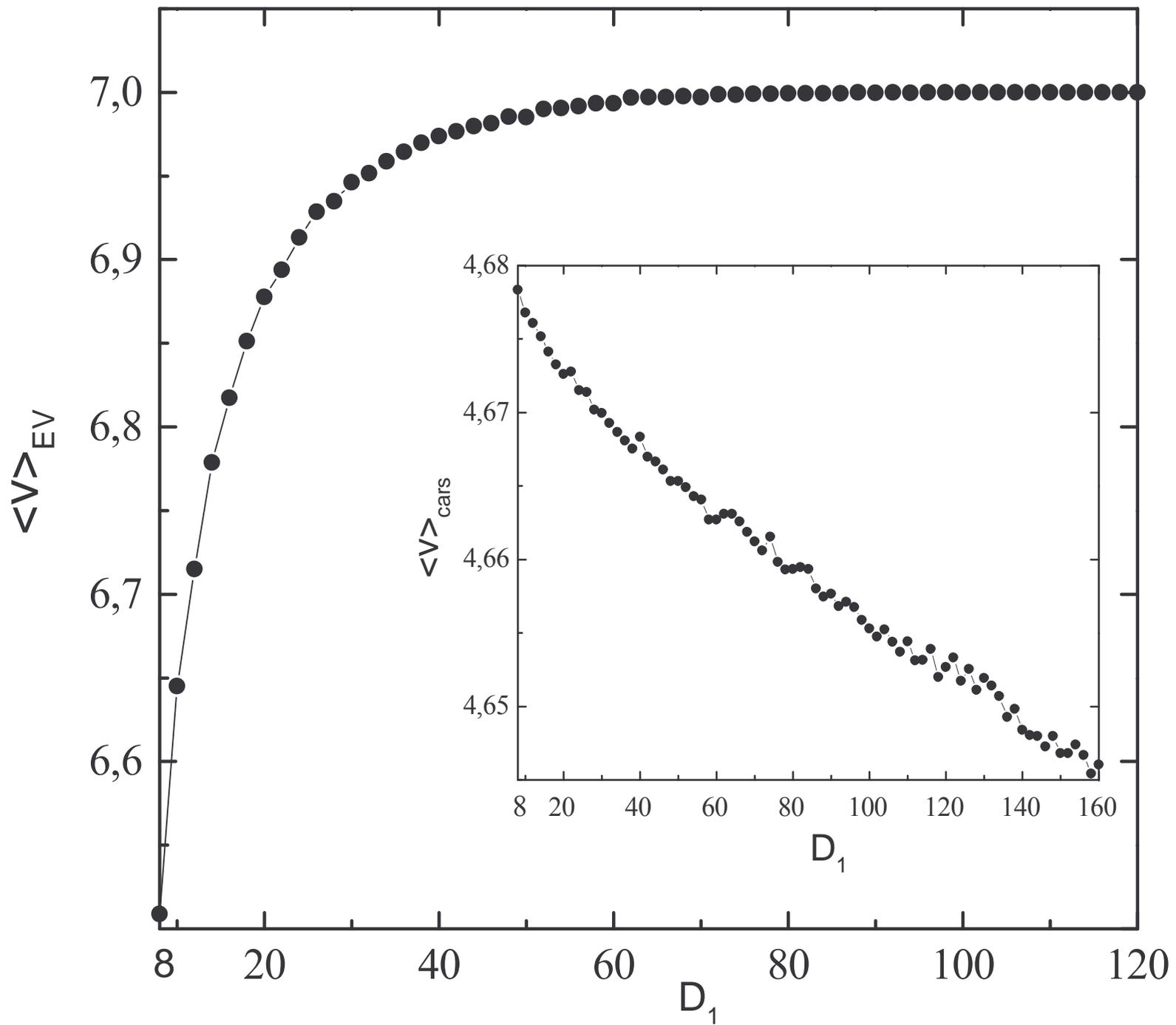
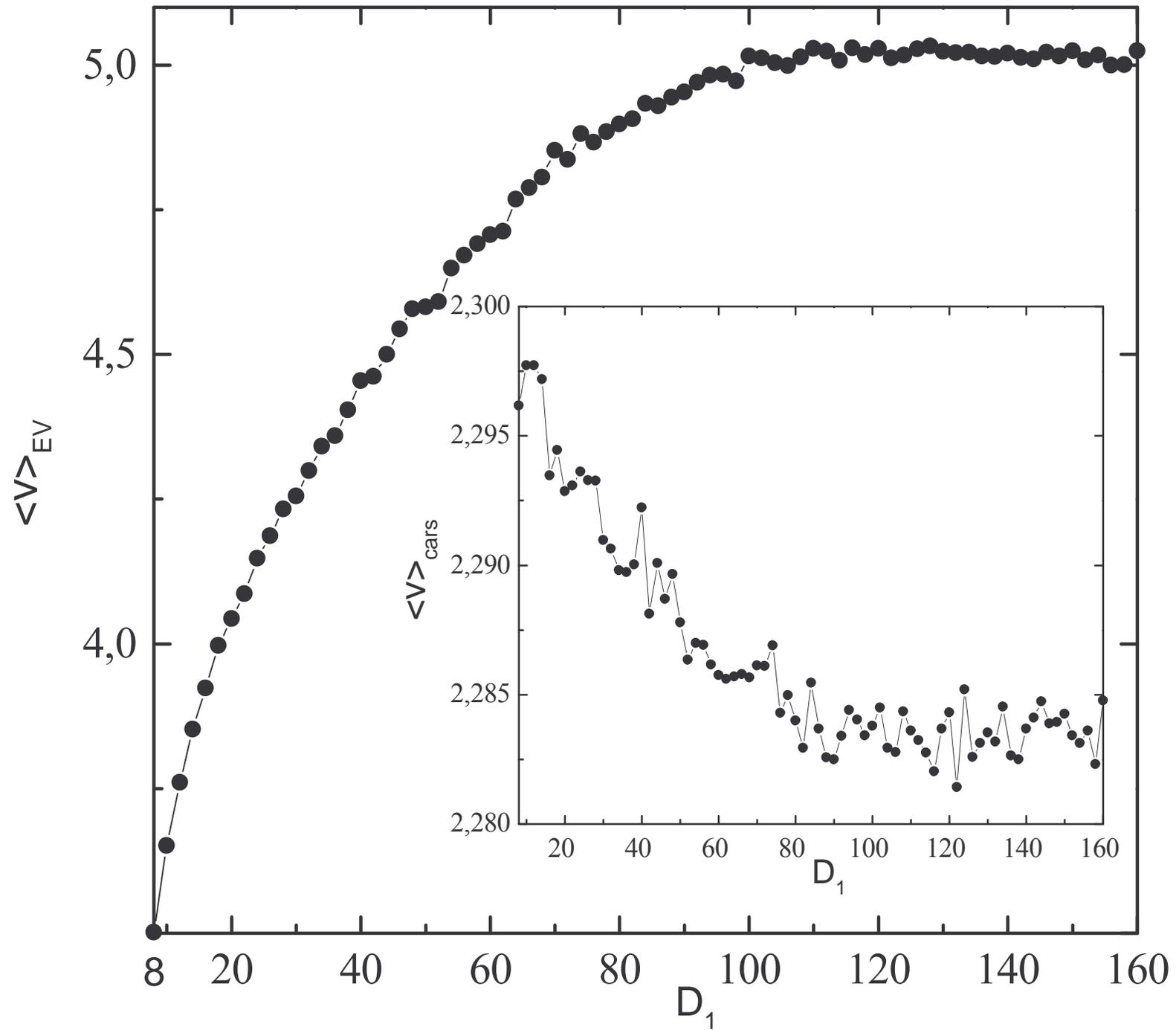
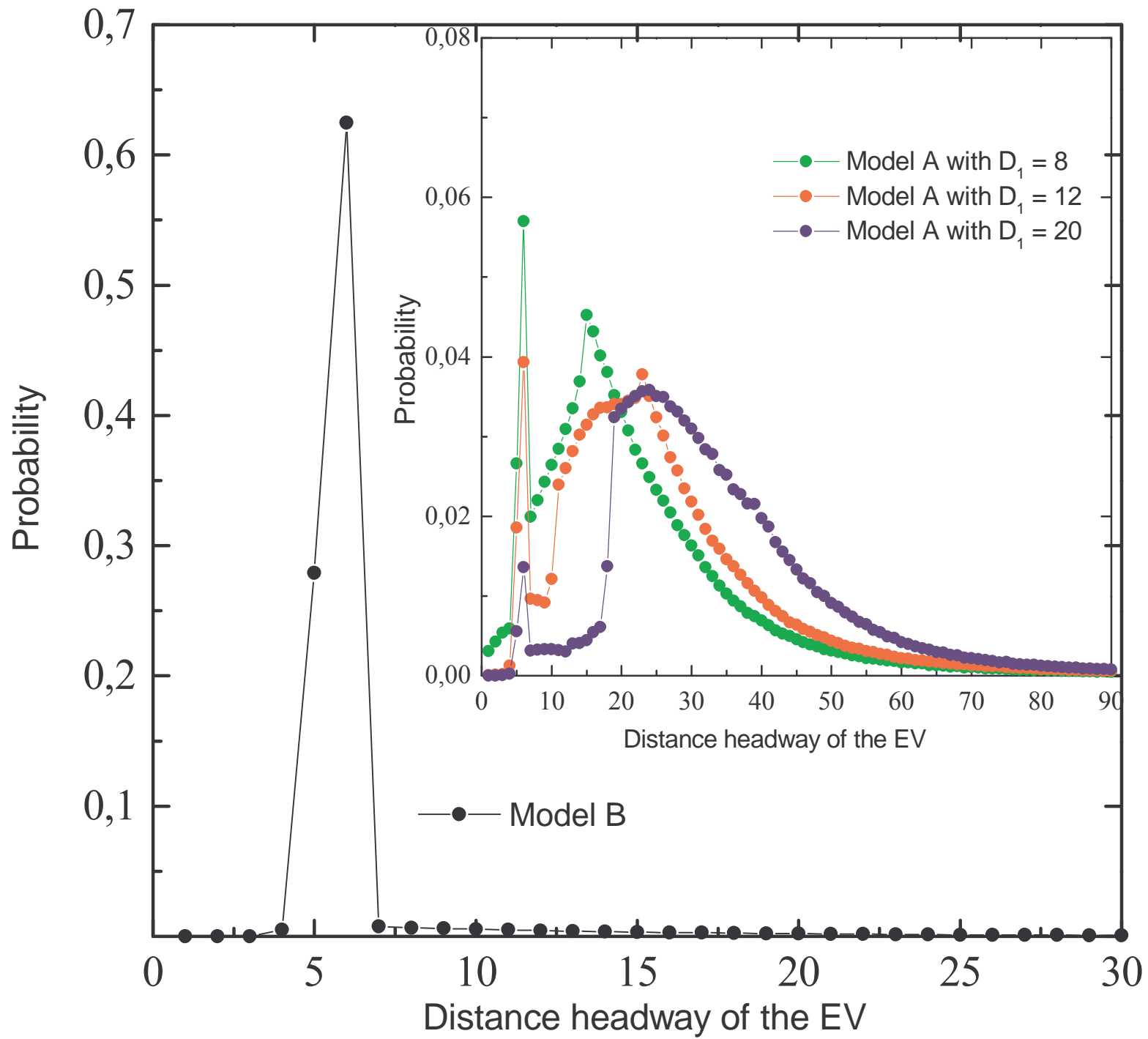
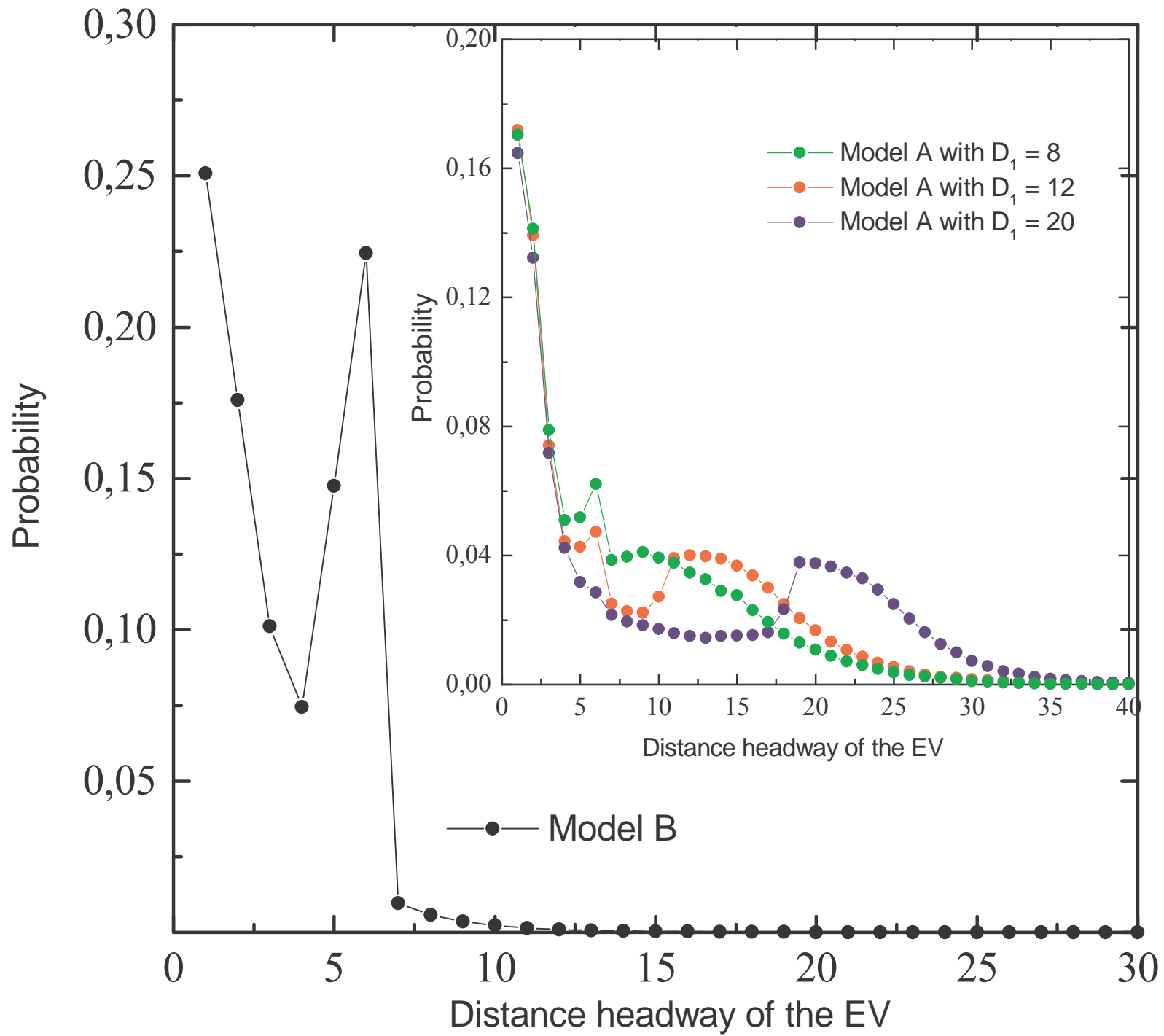
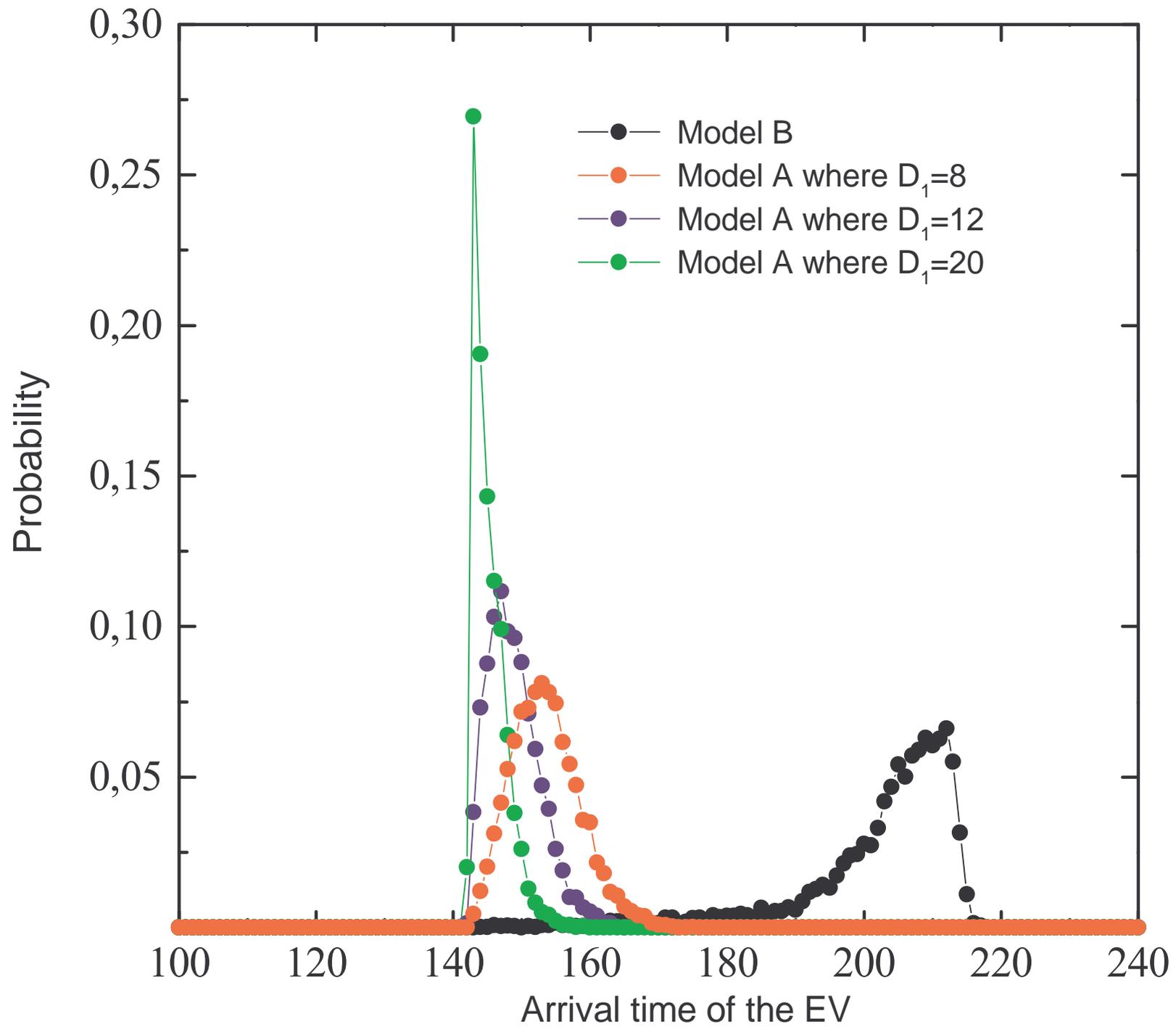
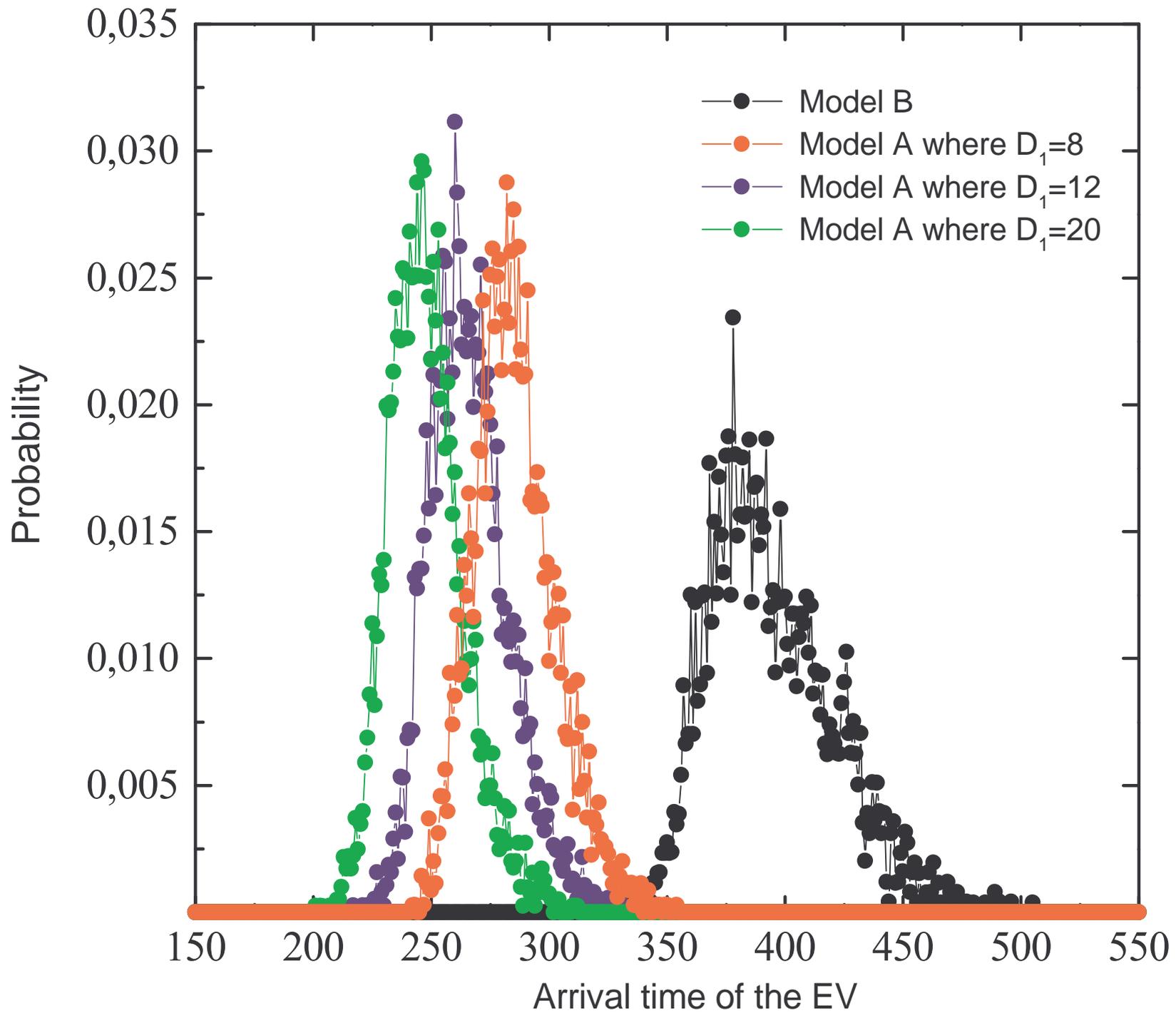
\end{document}